\documentclass[letterpaper, 10 pt, conference]{ieeeconf}

\IEEEoverridecommandlockouts                             
\usepackage{graphicx}
\usepackage{subcaption}
\usepackage{amsmath}
\usepackage{amssymb} 
\usepackage{caption}
\usepackage{booktabs}
\usepackage{algorithmicx}
\usepackage{algorithm}
\usepackage{algpseudocode}
\def\biblio{\bibliographystyle{IEEEtran}\bibliography{library.bib}}
\title{An Unsupervised Homogenization Pipeline for Clustering Similar
  Patients using Electronic Health Record Data}

\author{Alvaro Ulloa, Anna Basile, Gregory J. Wehner, Linyuan Jing, Marylyn D. Ritchie, Brett Beaulieu-Jones,\\
  Christopher M. Haggerty, Brandon K. Fornwalt%
  \thanks{AU, LJ, MR, CH, and BF are with Geisinger, PA, USA. {\tt\small
      [aeulloacerna,ljing,mdritchie,cmhaggerty] @geisinger.edu, bkf@gaech.edu}}
  \thanks{AB is with the Pennsylvania State University, PA, USA. {\tt\small azo121@psu.edu}}
  \thanks{GW is with the U. of Kentucky, KY, USA.  {\tt\small wehnergj@gmail.com}}
  \thanks{BB is with Harvard School of Medicine, MA, USA. {\tt\small brett\_beaulieu-jones@hms.harvard.edu}}}

\begin{document}
\def\biblio{}
\maketitle
\thispagestyle{empty}
\pagestyle{empty}

\begin{abstract}
  Electronic health records (EHR) contain a large variety of
  information on the clinical history of patients such as vital signs,
  demographics, diagnostic codes and imaging data. The enormous
  potential for discovery in this rich dataset is hampered by its
  complexity and heterogeneity.

  We present the first study to assess unsupervised homogenization
  pipelines designed for EHR clustering. To identify the optimal
  pipeline, we tested accuracy on simulated data with varying amounts
  of redundancy, heterogeneity, and missingness. We identified two
  optimal pipelines: 1) Multiple Imputation by Chained Equations
  (MICE) combined with Local Linear Embedding; and 2) MICE, Z-scoring,
  and Deep Autoencoders.
\end{abstract}

\section{Introduction}
Doctors provide diagnoses to help predict the health trajectory of
their patients. A diagnosis also helps to predict what treatments have
the highest likelihood of improving a patient’s health. The more
granular the diagnosis, the more specific or ``precise'' medicine can
become. The wealth of medical data gathered from patients, that is
digitally available in an electronic health record (EHR), should
support highly granular diagnoses. Unfortunately, the current clinical
paradigm of a human physician wading through this vast sea of data
cannot deliver the promise of precision medicine.

Fortunately, advances in machine learning can be harnessed to sift
through this rich dataset and extract useful information to facilitate
human decisions. One popular application is phenotyping by cluster
analysis. Previous studies~\cite{guan2016unsupervised,
  shah2015phenomapping, katz2017phenomapping} have shown that
clustering algorithms have the potential to classify patients into
similar phenotypes based on data contained in the medical record. For
example, using unbiased hierarchical cluster analysis and penalized
model-based clustering, Shah et al.~\cite{shah2015phenomapping}
identified 3 phenotypes in patients diagnosed with heart failure with
preserved ejection fraction. Upon identification of such granular and
more homogeneous clusters, the outcomes (e.g. hospitalization, cardiac
events or mortality) and attempted therapies within each cluster can
then be linked together to predict likely outcomes resulting from
choosing particular therapies.

While there have been many advances in the field of cluster
analysis~\cite{jain2010data}, the methods rely on the assumption of
homogeneous, non-redundant and complete data. However, EHR data are
heterogeneous (variables can be continuous or categorical, and with
different scales), redundant (multiple measurements may assess the
same underlying patient feature), incomplete (many fields in the
clinical reports are sparsely filled depending on the purpose of the
study), and noisy (not all variables are informative in all
conditions). Additionally, human errors and system biases also
contribute to measurement errors in EHR data. Thus, to fully utilize
the EHR to reliably detect disease subtypes, clustering techniques
must be paired with pre-processing techniques that normalize and
reduce the complexity of the raw EHR data. Such a clustering pipeline,
including pre-processing steps, has not been previously proposed or
validated.

In this paper, we assess and propose an optimal clustering pipeline
that is robust to the nuisances of EHR data. The pipeline consists of
imputation, normalization, feature reduction, and clustering. Multiple
commonly used techniques are evaluated at each step, and the best
performing pipeline is selected. Since the accuracy of clusters in
real EHR applications cannot be measured due to lack of a ground
truth, we assessed accuracy using simulated EHR data where ground
truth could be easily defined. To the best of our knowledge, this is
the first study to propose and validate an unsupervised homogenization
pipeline for EHR clustering.

\section{Methods}

\subsection{EHR Data Simulation}
\label{sec:setup}
We simulated patient encounters with a sample generator that mimics
the redundancy and heterogeneity of EHR data. We defined rows for
patient encounters (samples) and columns for measurements taken from
the patient (features). We designed three clusters with $n$ samples
per cluster, observed dimensionality $m$, and effective dimensionality
of 2 (for visualization convenience).

The sample generator drew $3n$ independent samples from a
multivariate normal distribution with $\mu=[0,0],$ and
$\Sigma=\mathbf{I}_{2 \times 2}$ to form the matrix $N_{3n \times
  2}$. Then, we separated the clusters by shifting $n$ samples at a
time. The first $n$ samples stayed in the origin, while the next $n$ were
shifted by $[d,0]$, and the last $n$ were shifted by
$[\frac{d}{2}, \frac{\sqrt{3}}{2}d],$ forming an equilateral triangle
with a distance $d$ from each vertex.

We emulated redundancy by projecting the original feature
vector to a $m$-dimensional space:
$X_{3n \times m} = N_{3n \times 2}P_{2 \times m},$ where the elements
of the projection matrix, $P$, were drawn from a uniform distribution
in the range $(0,1).$

We then enforced heterogeneity by quantizing half of the variables (set
to zero if below the mean and 1 otherwise), chosen at random, and
scaling each continuous feature with a random factor between
1 and 100. Finally, we added Gaussian noise ($\mu=0, \sigma=1$) to
every element in the data matrix to mimic measurement errors.

\begin{figure} 
  \centering
  \includegraphics[width=\columnwidth]{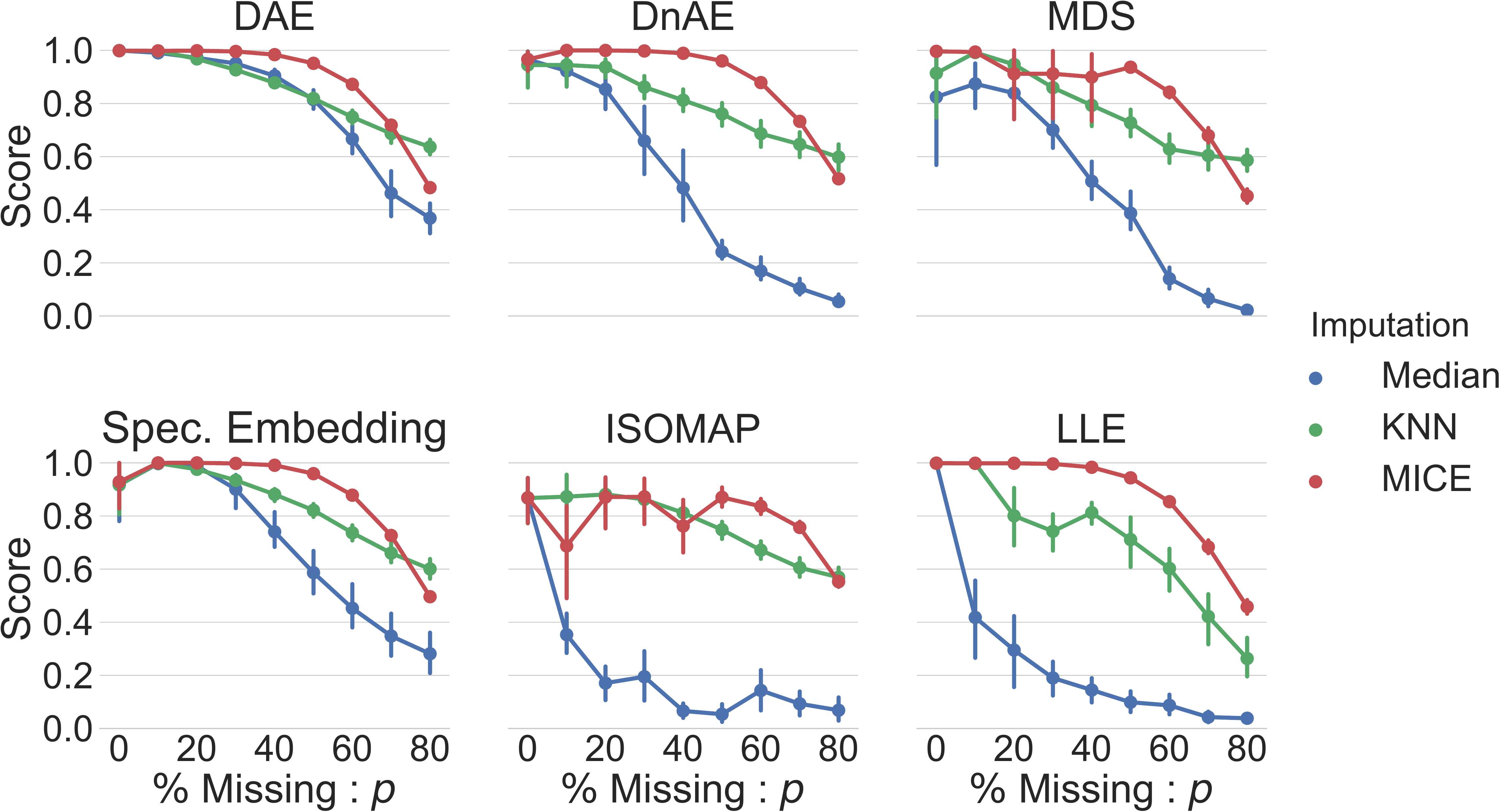}
  \caption{Missingness experiment results.}\label{fig:miss2}
\end{figure}
  
\subsection{EHR Clustering Pipeline}

\subsubsection{Imputation}
We tested median imputation, where the median value from valid samples
complete missing values; k-Nearest Neighbors (KNN), where the average
value from the k-nearest samples is used; and Multiple Imputation by
Chained Equations (MICE)~\cite{buuren2011mice}, where the missing
values are predicted based on regression models with complete samples.

\subsubsection{Normalization}
For continuous variables, we tested Z-score, where every variable is
set to zero mean and unit variance; MinMax, which normalizes to a
[0,1] range; and Whitening, where the feature space is linearly
projected such that inter-feature covariance is the identity matrix.

\subsubsection{Feature reduction}

We propose the use of Deep Autoencoders
(DAE)~\cite{hinton2006reducing} and Denoising Autoencoders
(DnAE)~\cite{vincent2008extracting} for EHR feature
reduction. Autoencoders are trained to reconstruct an input through
encoding and decoding networks. In the DnAE case, noise is added to
the encoded units to enforce robustness to measurement noise.

We designed the network architechture with a hyper-parameter search
for layers, hidden, and encoding units. The network with the least
number of encoding units that achieves the reconstruction error of 1\%
or less is preferred. The encoding vectors represent EHR data in a
compressed and continuous vector, suitable for any clustering
technique.

For comparison, we evaluated other methods with local (Local Linear
Embedding (LLE)~\cite{roweis2000nonlinear}) and global neighbor
algorithms (Isometric Mapping (ISOMAP)~\cite{tenenbaum2000global}); as
well as affinity matrix algorithms, such as Spectral
Embedding~\cite{ng2002spectral} and Multidimensional Scaling
(MDS)~\cite{borg2005modern}.

\subsubsection{Clustering}

Without loss of generality, we used K-means to conduct the final
cluster analysis.

\subsection{Simulation Setup and Experiments}

First, we simulated a baseline scenario where all parameters were set
to an ideal level with complete, free of noise, $d=10$, 5000 samples
per cluster, and $m=10$. An effect size of 10 resulted in less than
0.01\% overlap between clusters, and heuristically $m=10$ resulted in
good performance for all pipelines. This baseline was used to identify
the best performing pair of normalization and feature reduction
methods, which were then used in the rest of the experiments.

We then simulated four scenarios for testing the pipeline robustness
at various levels of severity. In all experiments, we swept one
simulation parameter while keeping all others constant.  We measured
the adjusted rand-score~\cite{hubert1985comparing}, which computes a
similarity measure between the results of two sets of labels by
counting pairs that are assigned in the same or different clusters in
the predicted and true clusterings while adjusting for random
chance. Table \ref{tab:experiments} describes each experimental setup
and the default parameters. Every experiment was run 5 times to
extract the mean and standard deviation of the performance.

\begin{table}
  \caption{Simulation experiments with default parameters $p=0$,
    $\eta=0$, $d=10$, $n=5000$, and $m=10$.}
  \label{tab:experiments}
  \centering
  \begin{tabular}{lrr}\toprule
    \textbf{Experiment}  & \textbf{Parameter} & \textbf{Levels} \\
    \midrule
    Effect Size & $d$       & [3, 4, 5, 6, 7, 8] \\
    Features    & $m$       & [6, 20, 40, 100, 200, 500]\\
    Missingness (\%) & $p$       & [0, 10, 20, $\dots,$ 80] \\
    Noise       & $\eta$    & [4, 16, 64, 128, 256]\\
    \bottomrule
  \end{tabular}
\end{table}

\subsubsection{Missingness}
To simulate the missing entries in the EHR, we randomly removed a
percentage $p$, from the observed data matrix and denoted them as
missing values. We varied $p$ from 0 to 80\% in increments of 10\%.

\subsubsection{Effect Size}
We manipulated the effect size by varying the distance between cluster
centers, $d$. In two dimensions, we can calculate the number of
overlapped samples by counting the number of samples beyond a distance
of $\frac{d}{2}$ in a bivariate standard normal distribution. Then, in
a triangular setting, the number of overlapped samples would be 6
times the calculated amount. By conducting a Monte Carlo simulation,
we can convert the effect sizes of 3--8 to the percentage of
overlapped samples [13.35\%, 4.55\%, 1.24\%, 0.27\%, 0.04\%,
0.01\%]. This can be interpreted as the lower-bound for error in
cluster assignment.

\subsubsection{Redundant Features}
We assessed the robustness to the number of redundant features present
in the dataset by increasing the dimensionality, $m$,
while keeping the ground-truth dimensionality of 2. We simulated
projection matrices that generated [6, 20, 40, 100, 200, 500]
features.

\subsubsection{Uninformative/Noisy Features}
EHR data contain information that may not be useful in determining
clusters of similar patients. We assessed the effects of including
non-informative variables by appending $\eta$ random continuous and
$\eta$ random binary variables.

\begin{table}
  \caption{Baseline results for identifying \textbf{best scaling for each
    feature reduction method}. The entries show average score and
    standard deviation of the scores across repetitions.}
  \label{tab:baseline}
  \centering
  \begin{tabular}{lllll}
\toprule
& \textbf{MinMax} & \textbf{Raw} & \textbf{Whitening} & \textbf{Z-score} \\
\midrule
\textbf{DAE        } &  0.982(0.03) &  0.822(0.20) &  0.841(0.15) &  \textbf{0.998(0.00)} \\
\textbf{DnAE       } &  0.983(0.03) &  0.769(0.20) &  0.781(0.20) &  \textbf{0.998(0.00)} \\
\textbf{MDS        } &  0.903(0.17) &  0.985(0.03) &  0.294(0.18) &  \textbf{0.999(0.00)} \\
\textbf{ISOMAP     } &  0.264(0.41) &  0.235(0.35) &  \textbf{0.822(0.26)} &  0.390(0.43) \\
\textbf{LLE        } &  0.737(0.20) &  \textbf{0.976(0.05)} &  0.503(0.32) &  0.745(0.21) \\
\textbf{Spec. Emb.} &  0.770(0.21) &  \textbf{0.994(0.01)} &  0.634(0.29) &  0.753(0.23) \\
\bottomrule
  \end{tabular}
\end{table}

\section{Results}

The baseline experiment revealed that the performance of the
clustering pipeline heavily depended on the choice of normalization
and feature reduction method (see Table~\ref{tab:baseline}).DAE, DnAE,
and MDS paired best with Z-scoring, all with scores above 0.99. ISOMAP
performed best with Whitening while LLE and Spectral Embedding
obtained its best performance when no scaling was used. We used these
optimal pairs to conduct the remainder of the experiments.

\subsection{Robustness Experiments}

\subsubsection{Missingness}
As shown in Fig.~\ref{fig:miss2}, levels of missingness above 60\%
significantly impaired the clustering performance for all pipeline
configurations (all scores below 0.8). Among the three imputation
methods, MICE resulted in the best performance for all feature
reduction methods except ISOMAP, for which KNN was marginally better
up to 50\%.  Median imputation consistently the worst performance.

\subsubsection{Effect Size}
As expected, the performance of all configurations increased with the
effect size (Fig. \ref{fig:effect}). Overall, the top three performing
feature reduction methods were LLE, DAE, and MDS. LLE exhibited the
best performance across feature reduction methods but only marginally
better than DAE, e.g. the p-value of a paired t-test was 0.03 at the
effect size of 4.

\subsubsection{Features}
LLE, DAE, and MDS were essentially immune to large amounts of
redundant features (Fig. \ref{fig:features}). DnAE appeared to be
similarly immune at low levels, but its performance sharply decreased
with greater than 200 features. Conversely, Spectral Embedding
benefited from higher numbers of redundant features and performed on
par to the best methods for 200 and 500 redundant features. ISOMAP
performed poorly at all levels

\subsubsection{Uninformative/Noisy Features}
\begin{figure}[t]
  \centering
   \subcaptionbox{Effect size $d$\label{fig:effect}}{\includegraphics[height=5.5\baselineskip]{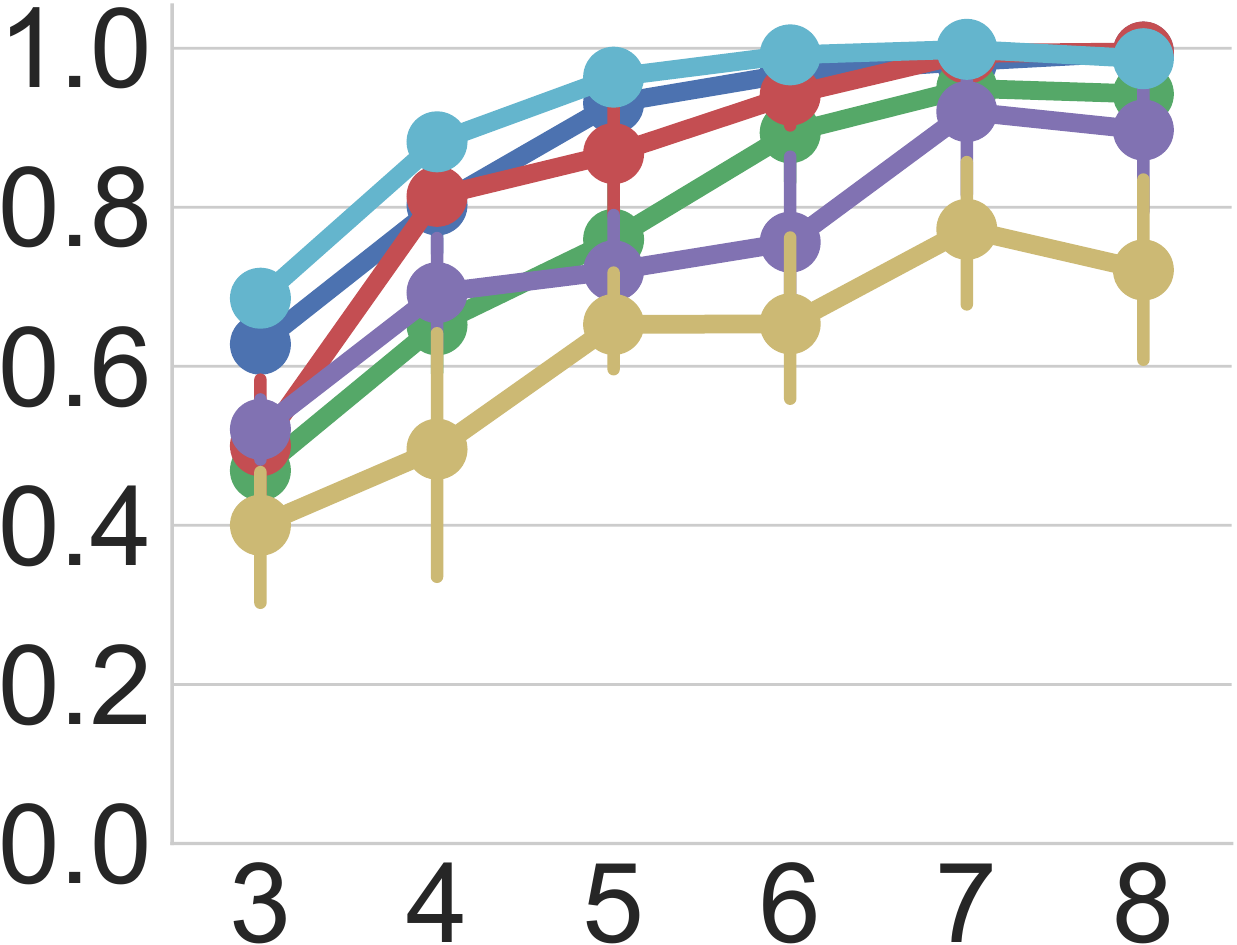}}%
   \subcaptionbox{Features $m$\label{fig:features}}{\includegraphics[height=5.5\baselineskip]{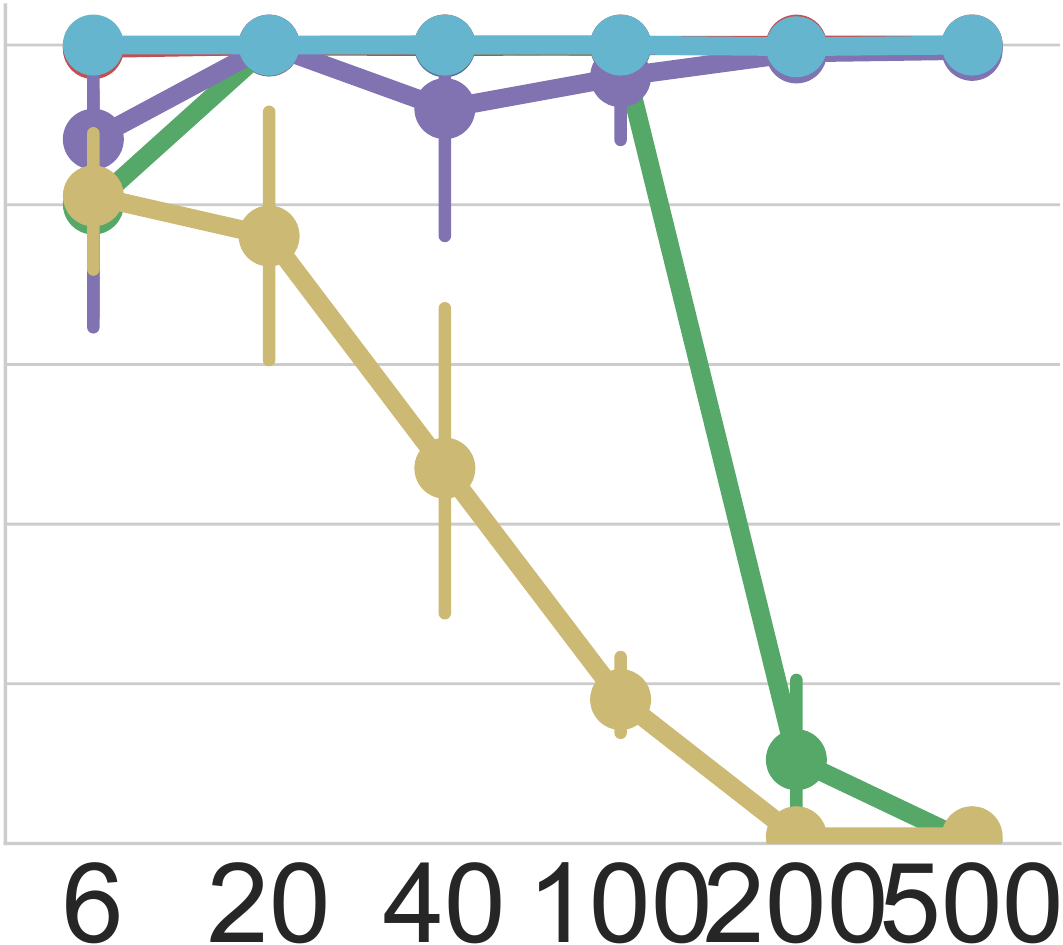}}%
   \subcaptionbox{Noise $\eta$\label{fig:noise}}{\includegraphics[height=5.5\baselineskip]{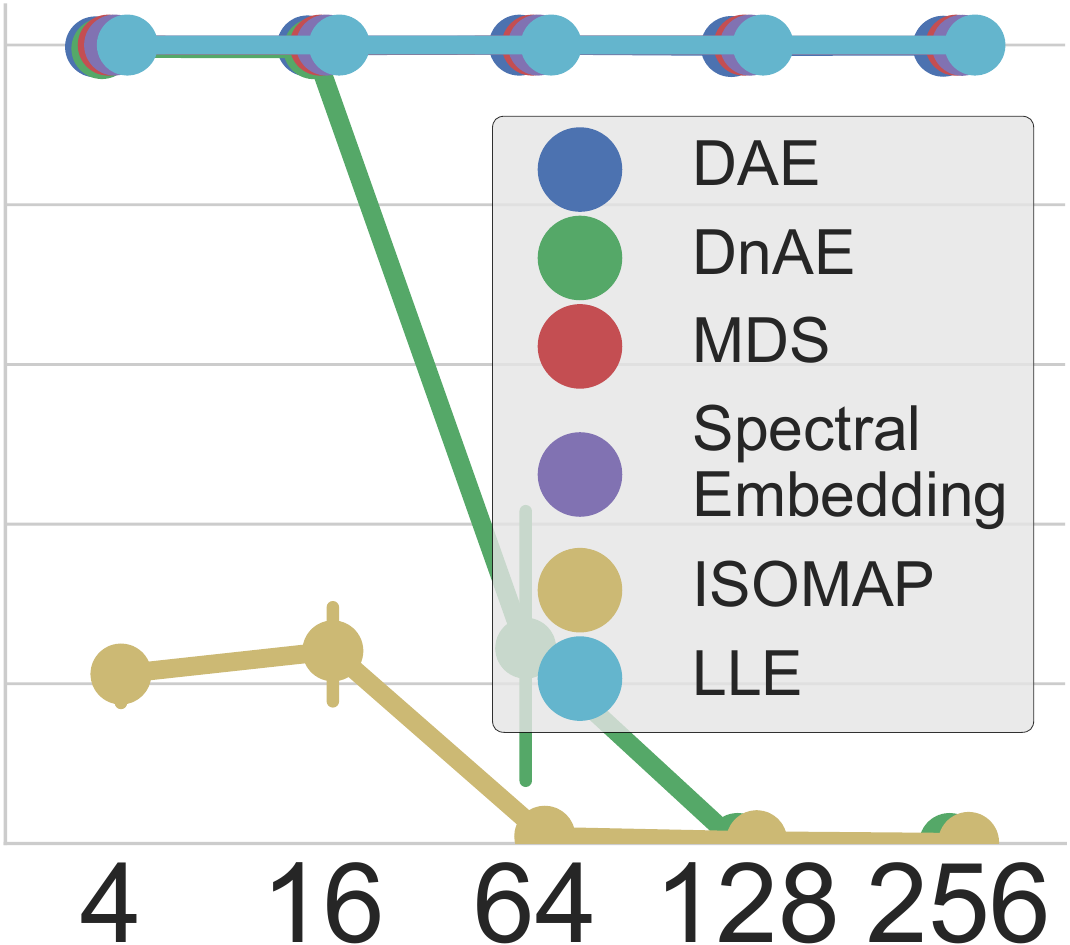}}%
  \caption{Robustness experiments results}\label{fig:robustness}
\end{figure}

As shown in Fig. \ref{fig:noise}, most methods, except DnAE and
ISOMAP, were immune to large amounts of uninformative variables. DnAE
was robust to uninformative variables up to 32 continuous and binary
uninformative variables. ISOMAP did not tolerate even the minimum
number of uninformative variables.

\subsection{Interaction Experiments}

Following the robustness experiments, we identified DAE and LLE as the
top 2 best performing feature reduction methods overall. To further
compare these methods, we performed subsequent experiments that
allowed for interactions of varying effect sizes, missingness, and
noise.

\begin{figure}
  \centering
  \subcaptionbox{Effect size vs Noise\label{fig:effect_noise}}{\includegraphics[height=9\baselineskip]{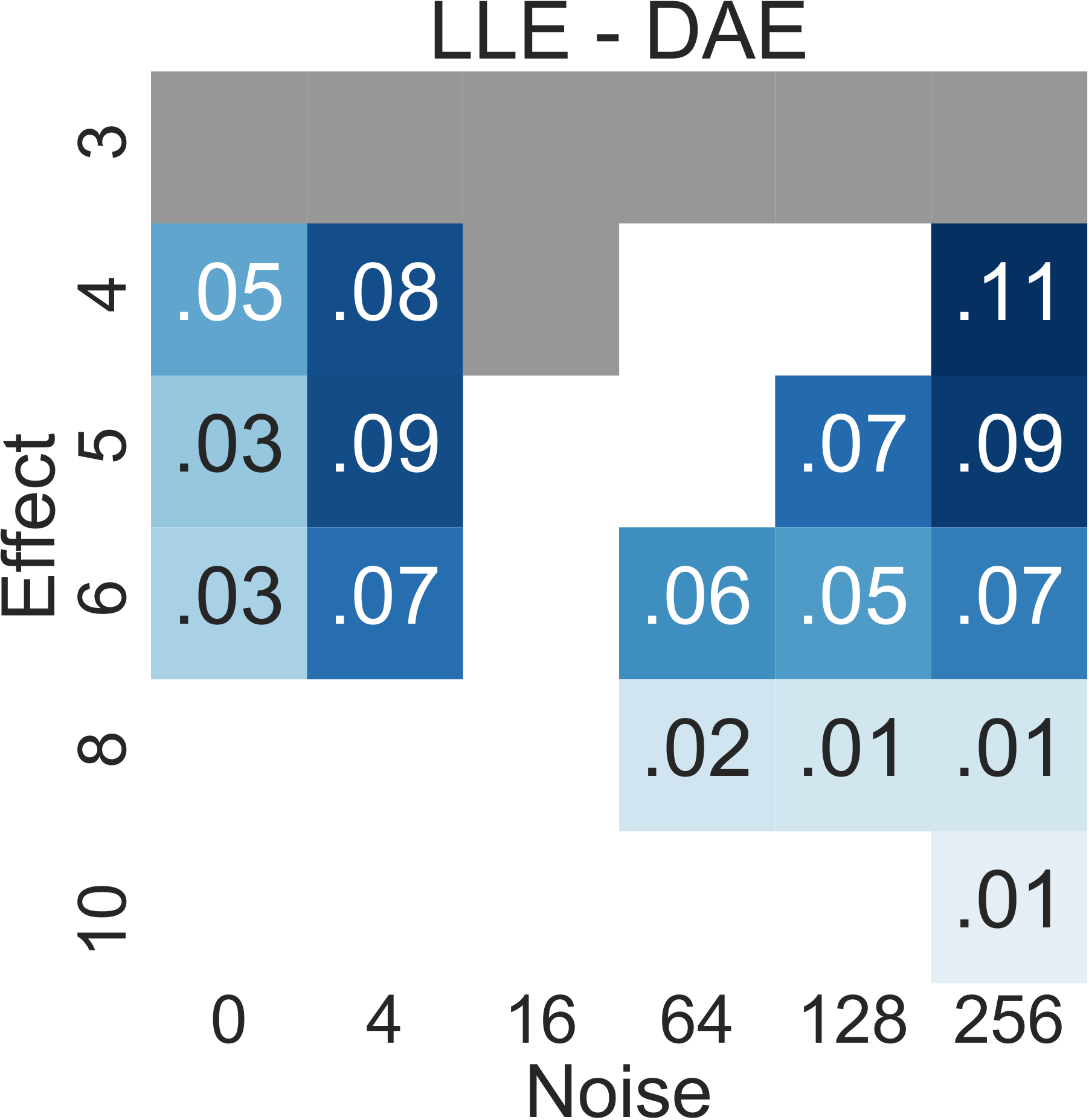}} %
  \subcaptionbox{Effect size vs Missingness\label{fig:effect_missing}}{\includegraphics[height=9\baselineskip]{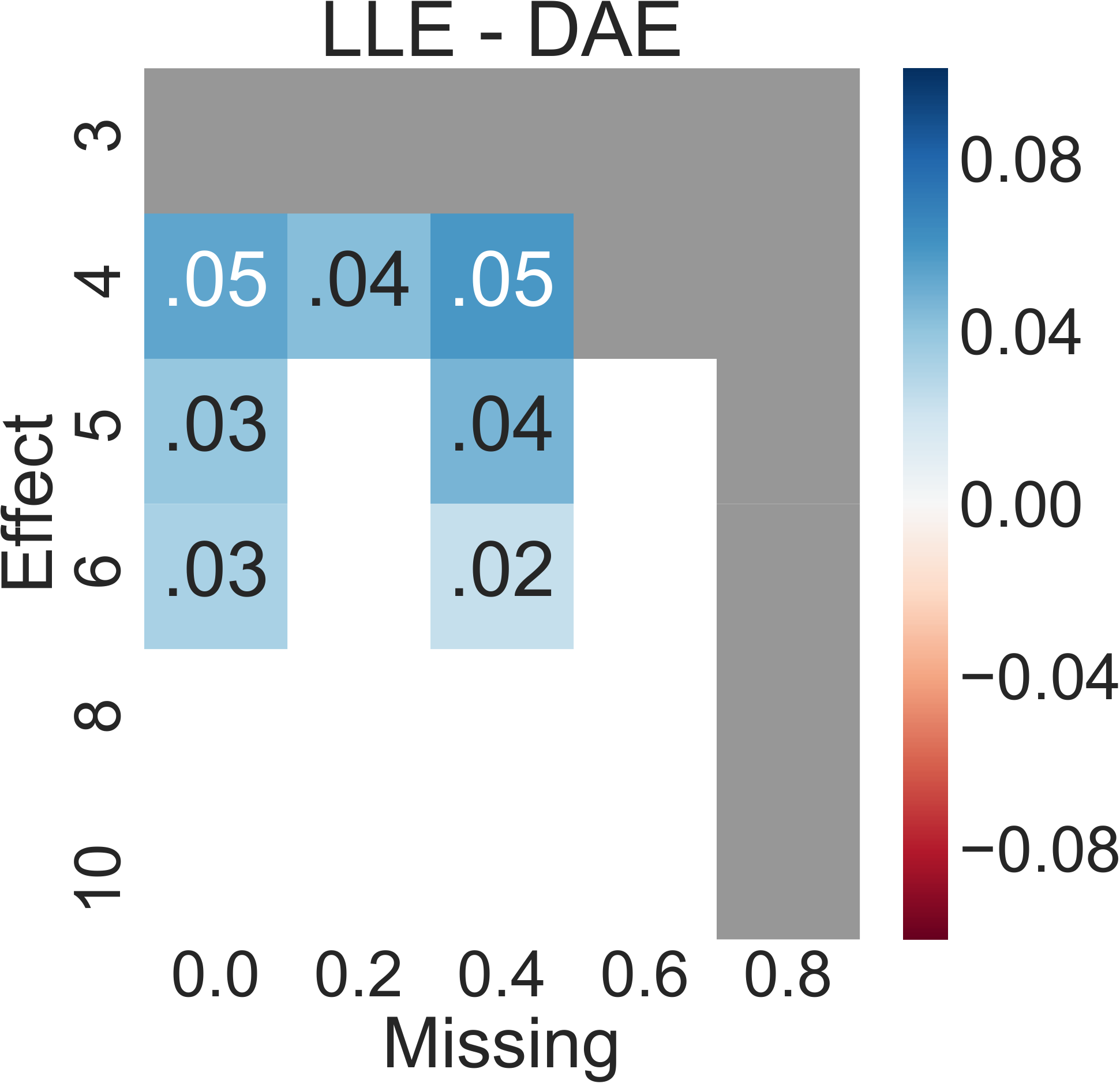}}%

  \caption{Interaction experiments. The gray areas denote
    where neither method scored above 0.8,
    the white areas denote no significant difference between score
    means. The colored areas denote significant differences between
    LLE and DAE.}\label{fig:interaction}
\end{figure}

Overall, LLE matched or outperformed DAE. In the effect size vs
noise experiment, (Fig.  \ref{fig:effect_noise}) large amounts of
uninformative variables and medium effect sizes favored LLE. In the
effect size vs missingness experiment, (Fig. \ref{fig:effect_missing})
LLE showed significantly better performance for medium effect sizes
and low missingness and no difference for large effect sizes and low
to medium missingness.

\section{Discussion}

In the current paper, we propose and evaluate a clustering pipeline
tailored for complex EHR data by comparing performances of commonly
used techniques. We found two pipelines that outperform other
alternatives: 1) MICE imputation + LLE feature reduction; 2) MICE
imputation + Z-score normalization + DAE feature reduction. Both
pipelines are robust to missingness (up to 60\%), uninformative noise
and large numbers of redundant features, while LLE performs slightly
better at smaller effect size. This is the first study to present an
unsupervised homogenization pipeline designed for EHR clustering.

\subsubsection*{Normalization}
EHR data are heterogeneous, containing both categorical and continuous
variables at different scales. Normalization is recommended to reduce
the variance among variables. Most previous
studies~\cite{shah2015phenomapping,miotto2016deep,beaulieu2016semi}
normalized EHR variables to a range of (0, 1), however, as shown in
Table~\ref{tab:baseline}, the best normalization method is closely
related to the feature reduction method. For example, for DAE and
DnAE, Z-score normalization results in the best performing pipelines,
while no normalization is necessary for LLE. This is reasonable since,
unlike DAE and other distance-based algorithms, neighbor-based
algorithms, such as LLE, eliminate the need to estimate distance
between objects.

\subsubsection*{Imputation}
Given the wide array of measurements that can be obtained from
patients, missing data are common, and it is impossible that every
patient has every possible test and measurement. Physicians evaluate
the cost-benefit of each test and may not request a particular test if
the result will not be informative for the diagnosis or treatment. We
evaluated a spectrum of imputation techniques that could induce
different levels of artificial similarity. The simulation results
favored MICE for all feature reduction methods except
ISOMAP. Consistent with our studies, MICE has also shown good
performance for life-history/EHR datasets in previous
studies~\cite{penone2014imputation,beaulieu2017characterizing}.

The main assumptions of MICE are that other non-missing values are
predictive of the missing ones (redundancy) and that the data are
missing-at-random. EHR data satisfies the redundancy assumption, for
example, age, sex, and height are known to be good predictors of
aortic root diameter~\cite{devereux2012normal}. White et
al.~\cite{white2011multiple} note that MICE is sensitive to departures
from the missing-at-random assumption. However, such assumptions can
be relaxed as long as the dataset contains enough complete samples to
build reliable predictive models. Theoretically, EHR data is likely to
follow a missing not at random over a missing at random mechanism, as
there is likely a reason for missing values (e.g. patient’s health,
physician’s recommendation, socioeconomic status). However, the true
pattern of missingness is likely influenced by both MAR and
MNAR. Hence, MICE can still be applied given an abundance of data.

\subsubsection*{Feature Reduction}
The EHR contains many redundant pieces of information. For example,
body mass index can be easily computed from height and weight. Thus,
it is necessary to reduce the redundancy to extract effective (and
possibly latent) features from this high dimensional dataset. Our
simulation results show that among the different feature reduction
methods, pipelines with DAE and LLE show the highest
accuracy. Moreover, LLE outperforms DAE by 0.05-0.12 at medium effect
size and high uninformative noise. This suggests that LLE might be
better at detecting granular phenotypes that have more overlapped
samples (1–-5\%, corresponding to an effect size of 4–-5). Additionally,
another benefit of using LLE is that no normalization to input data is
needed, as discussed above.

However, compared to LLE, DAE is more computationally efficient, that
is $O(m \log(k) n \log(n))$ vs $O(nm)$, where $k$ denotes the number
of neighbors for LLE. Once the network is trained, the weights can be
applied to a new dataset with minimal computation, while LLE computes
and sorts distances to all neighbors. Thus, considering the
large-scale nature of the EHR data, DAE might be a better choice when
used to make predictions for future patients. Recent studies deep
auto-encoders have demonstrated their ability to identify meaningful
representations of EHR
data~\cite{miotto2016deep,beaulieu2016semi}. Miotto et al.\ first
proposed the use of deep autoencoders for EHR data and called its
representation ``Deep patient''~\cite{miotto2016deep}. They
demonstrated its utility by assessing the probability of patients
developing various diseases and showing improvement in classification
scores for 76,214 patients and 78 different diseases. Similarly,
Beaulieu-Jones et al.\ reported improved classification scores for
amyotrophic lateral sclerosis diagnosis in clinical trials using
10,723 patients~\cite{beaulieu2016semi}. These are promising results
which demonstrate the potential of the proposed pipeline with DAE to
utilize EHR data to identify granular disease phenotypes, and to
ultimately facilitate precise diagnoses, risk prediction and treatment
strategies. Moreover, while these previous studies have shown the
promise of DAE, this is the first study to validate and design the
entire pipeline for clustering.

\subsubsection{Conclusions}
In conclusion, we propose an unsupervised homogenization pipeline to
fully integrate all components of EHR data for clustering
patients. After MICE imputation, both LLE with raw features and DAE
with z-score normalization show good clustering results. While LLE
marginally outperformed DAE in several direct comparisons, the
computational efficiency of DAE in evaluating new observations based
on large- scale EHR data (as is desired for precision medicine
approaches) provides an important advantage. Future studies are
required to evaluate and compare the two pipelines in real clinical
scenarios with large-scale EHR data.

\section*{Acknowledgement}
This project was also funded, in part, under a grant with the
Pennsylvania Department of Health (\#SAP 4100070267).

\bibliographystyle{IEEEtran}
\bibliography{library}
\end{document}